\documentclass[sigconf,natbib=false]{acmart}
\usepackage{graphicx} 
\usepackage{subcaption} 
\usepackage{wrapfig}
\usepackage{dblfloatfix} 

\usepackage{booktabs}   
\usepackage{multirow}   
\usepackage{makecell}   
\usepackage{threeparttable}

\usepackage[normalem]{ulem}
\usepackage{microtype}
\usepackage{algorithm}

\usepackage[table]{xcolor}
\usepackage[most]{tcolorbox}

\usepackage{pgfplots}
\pgfplotsset{compat=1.18}
\usepackage{tikz}

\usepackage{hyperref}
\usepackage{url}

\definecolor{aframe}{RGB}{144,188,232} 
\definecolor{aback}{RGB}{248,248,248} 
\newtcolorbox{questionbox}[1][]{
enhanced,
colback=aback,
colframe=aframe,
boxrule=1.3pt, 
arc=4mm, 
left=6mm,
right=6mm,
top=3.5mm,
bottom=3.5mm, 
boxsep=0mm,
width=\linewidth, 
center, 
#1
}

\definecolor{ref_col}{RGB}{184,96,42}
\definecolor{ref_col2}{RGB}{191,83,13}

\definecolor{ref_col}{RGB}{184,96,42}
\definecolor{ref_col2}{RGB}{191,83,13}
\definecolor{longc}{RGB}{73,88,187} 
\definecolor{shortc}{RGB}{73,88,187} 

\definecolor{cibg}{RGB}{242,242,242}
\definecolor{cifg}{RGB}{90,90,90}

\newtcbox{\cibadge}{on line,
colback=cibg,
colframe=cibg,
boxrule=0pt,
arc=2pt,
left=2pt,right=2pt,top=1pt,bottom=1pt,
boxsep=0pt
}

\copyrightyear{2026}
\acmYear{2026}
\setcopyright{cc}
\setcctype{by}
\acmConference[KDD 2026]{Proceedings of the 32nd ACM SIGKDD Conference on Knowledge Discovery and Data Mining V.2}{August 9--13, 2026}{Jeju Island, Republic of Korea}
\acmBooktitle{Proceedings of the 32nd ACM SIGKDD Conference on Knowledge Discovery and Data Mining V.2 (KDD 2026), August 9--13, 2026, Jeju Island, Republic of Korea}
\acmDOI{10.1145/3770855.3818881}
\acmISBN{979-8-4007-2259-2/2026/08}
\RequirePackage[
  datamodel=acmdatamodel,
  style=acmnumeric,
  doi=false,
  url=false,
  isbn=false,
  ]{biblatex}
\AtBeginBibliography{%
  \sloppy
  \raggedright
  \setlength{\emergencystretch}{6em}%
}

\AtBeginDocument{
  \setlength{\textfloatsep}{4pt plus 2pt minus 2pt}
  \setlength{\floatsep}{4pt plus 2pt minus 2pt}
  \setlength{\intextsep}{4pt plus 2pt minus 2pt}

  \setlength{\dbltextfloatsep}{6pt plus 2pt minus 2pt}
  \setlength{\dblfloatsep}{4pt plus 2pt minus 2pt}

  \setlength{\abovecaptionskip}{3pt}
  \setlength{\belowcaptionskip}{0pt}
}

\begin{document}

\title{CAP: Towards PPG Universal Representation Learning with Patient-level Supervision}

\author{Chenyang He}
\authornote{Both authors contributed equally to this research.}
\affiliation{%
  \institution{Nanjing University of Aeronautics and Astronautics}
  \city{Nanjing}
  \state{Jiangsu}
  \country{China}
}
\email{raoquan00@gmail.com}

\author{Xinyi Shao}
\authornotemark[1]
\affiliation{%
  \institution{Nanjing University of Aeronautics and Astronautics}
  \city{Nanjing}
  \state{Jiangsu}
  \country{China}
}
\email{xshao@arizona.edu}

\author{Shun Huang}
\affiliation{%
  \institution{Peking University}
  \city{Beijing}
  \state{Beijing}
  \country{China}
}
\email{huangshun0815@gmail.com}

\author{Bosong Huang}
\affiliation{%
  \institution{Independent Researcher}
  \country{N/A}
}
\email{boson.hwang@gmail.com}

\author{Daoqiang Zhang}
\affiliation{%
  \institution{Nanjing University of Aeronautics and Astronautics}
  \city{Nanjing}
  \state{Jiangsu}
  \country{China}
}
\email{dqzhang@nuaa.edu.cn}

\author{Ming Jing}
\affiliation{%
  \institution{Independent Researcher}
  \country{N/A}
}
\email{mingjinedu@gmail.com}

\author{Cheng Ding}
\authornote{Corresponding author: chengding@nuaa.edu.cn}
\affiliation{%
  \institution{Jinling Clinical Medical College}
  \institution{College of Artificial Intelligence}
  \institution{Nanjing University of Aeronautics and Astronautics}
  \city{Nanjing}
  \state{Jiangsu}
  \country{China}
}
\email{chengding@nuaa.edu.cn}

\renewcommand{\shortauthors}{He et al.}

\begin{abstract}
Photoplethysmography (PPG) plays a central role in wearable health monitoring and clinical decision support. Yet existing approaches to universal PPG representation learning largely focus on signal-level objectives and often overlook patient-level health context, which limits generalization to complex clinical tasks and heterogeneous cohorts. To address this gap, we construct a large-scale paired PPG-EHR multimodal dataset by distilling fragmented medical histories and clinical records into cohesive, patient-level electronic health records (EHR). Building on this resource, we propose \textbf{C}linical \textbf{A}nchored \textbf{P}retraining for PPG (\textbf{CAP}). During pretraining, CAP performs cross-modal contrastive alignment that anchors PPG representations to patient-level clinical semantics, guiding the encoder beyond waveform fitting toward modeling consistency in a patient’s overall physiological state. During downstream adaptation, the pretrained PPG encoder provides clinically grounded representations that strengthen inductive bias and improve robustness and transferability. Experiments demonstrate that CAP consistently outperforms strong baselines on four diverse downstream tasks. CAP achieves a particularly large gain on respiratory rate prediction (up to \(\mathbf{+87.6\%}\) relative improvement over the state-of-the-art baseline) and delivers an average relative \(\mathbf{+26.7\%}\) across all tasks. We further enhance the interpretability of our approach through comprehensive analyses, including ablations and multiple complementary visualizations of the learned representations. The code for our experiments is available at: \url{https://github.com/gody123gody/CAP}.

\end{abstract}



\maketitle

\section{Introduction}
\begin{figure}[h]
  \centering
  \includegraphics[width=\linewidth]{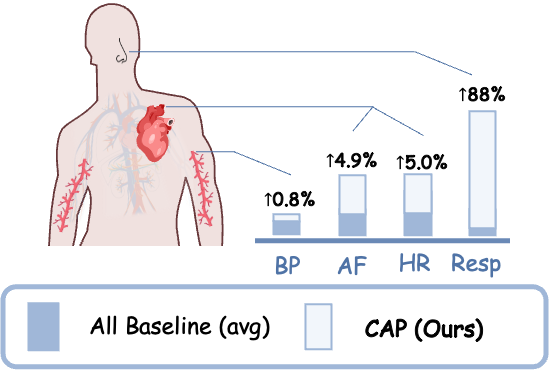}
  \caption{Overall performance of CAP across downstream tasks, contains Blood pressure (BP) estimation, Atrial fibrillation (AF) detection, Heart rate (HR) estimation and Respiratory rate (Resp) estimation.}
  \label{Intro_c1}
\end{figure}

Photoplethysmography (PPG) is a low-cost, non-invasive physiological~\cite{charlton20232023,jiang2025ppg,breen2025advancing} sensing modality that has been widely deployed in mobile health (mHealth)~\cite{li2025iot} and clinical decision support~\cite{wang2025thinking}, enabling key applications such as heart rate and respiratory monitoring~\cite{miao2025respdiff}, blood pressure estimation~\cite{botrugno2025optimized}, and arrhythmia screening~\cite{chen2025gptppg,ding2024siamquality}. Learning a transferable Universal PPG Encoder that captures robust physiological representations~\cite{fang2025ppgflowecg,fang2026ecgflowcmr} has become an important direction in medical informatics, which can be reused across diverse downstream tasks.

However, existing PPG studies largely focus on deriving medical inductive patterns from short term PPG segments, while overlooking the patient’s underlying physiological state reflected by the signal~\cite{xue2025ppg,lee2025cooperative,chen2025gptppg}. We therefore claim that:

\begin{questionbox}
\itshape
PPG representation learning should be patient-centric.
\end{questionbox}

\begin{figure}[h]
  \centering
  \includegraphics[width=\linewidth]{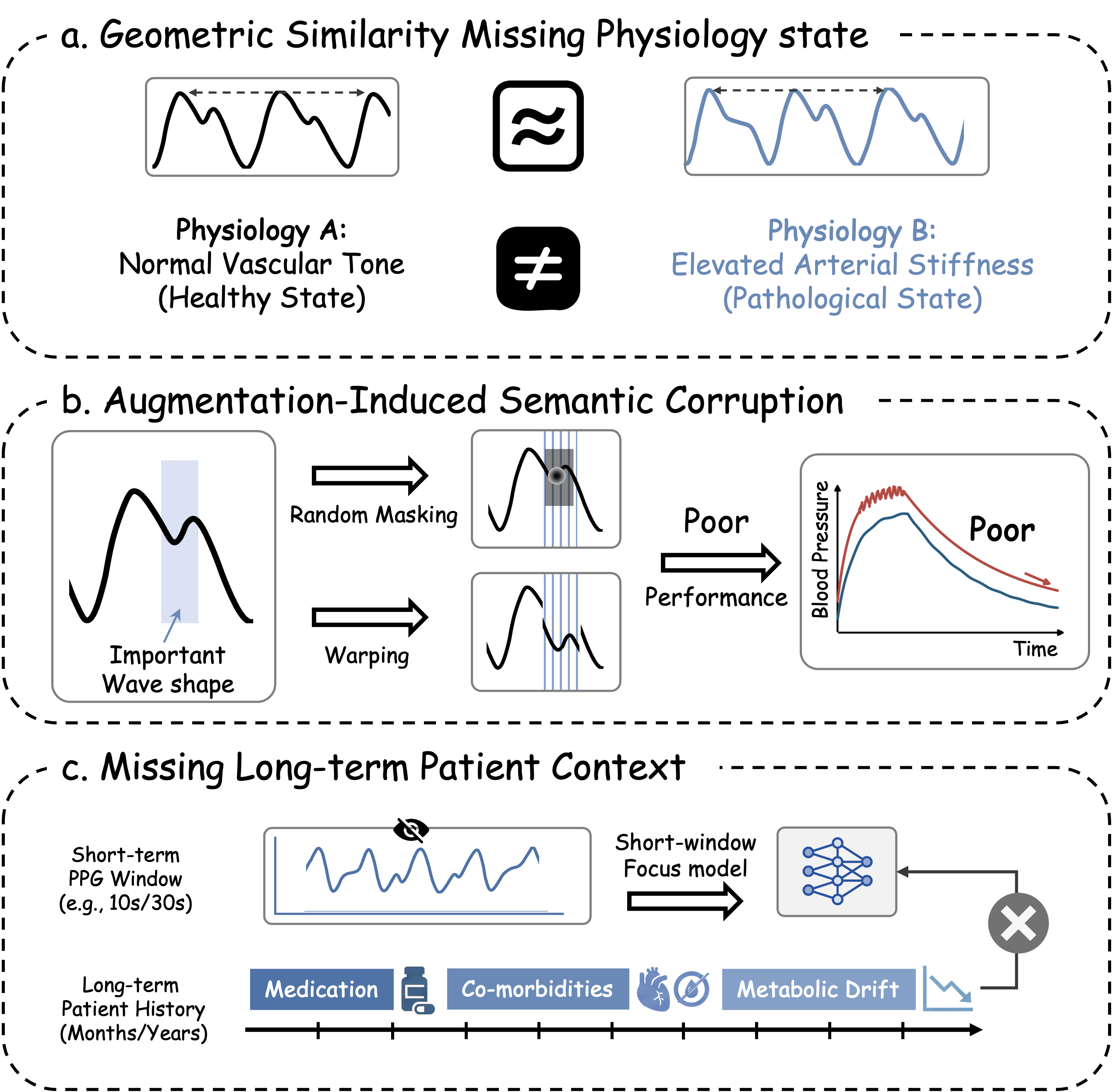}
  \caption{Challenges of pSSL (PPG Self-Supervised Learning, referring to approaches that rely solely on PPG signals themselves for representation learning). (A) Over-reliance on geometric waveform similarity. (B) Semantic corruption introduced by common augmentations. (C) Limited temporal scope.}
  \label{Intro_c2}
\end{figure}

These limitations are illustrated in Figure~\ref{Intro_c2}. First, semantic sparsity and bias in the representation space~\cite{chen2020simple}. Many contrastive approaches primarily align samples by geometric waveform similarity~\cite{ma2026self}. Yet PPG is an external manifestation of complex feedback within the circulatory system, and similar waveforms do not necessarily imply the same physiological or pathological state. This mismatch limits the model’s ability to distinguish signals that are morphologically similar but semantically different.

Second, augmentation-induced semantic corruption. Common augmentations such as masking, temporal perturbations, or morphological transforms~\cite{liang2023shapelet,he2025pefnet} may inadvertently distort fine-grained structures that carry clinically relevant information, thereby constraining transferability to fine-grained diagnosis or sensitive estimation tasks.

Third, limited temporal-scale awareness. Many studies pretrain on short segments (e.g., 5--30 seconds)~\cite{ding2024siamquality,ismail2022heart}, whereas real-world clinical and out-of-clinic monitoring is tightly coupled with a patient’s long term context. Relying solely on short windows often fails to capture chronic conditions and long-horizon physiological drift, which in turn undermines generalization across tasks and cohorts.

To address these challenges, we first expand the temporal field of view by modeling longer PPG segments, enabling the encoder to capture long-range physiological dynamics. We then incorporate patient-level clinical history within the corresponding time window as semantic supervision, anchoring representation learning to clinically meaningful patient states rather than waveform-level heuristics. Together, these designs promote patient-centric representations.

To support this paradigm, we construct a large-scale dataset with \textbf{2,279} patients. For each patient, we pair long term PPG signals with an electronic health records (EHR) distilled by a pretrained language model (PLM) from fragmented diagnosis codes and narrative clinical histories, providing rich patient-level clinical context as supervision for the paired PPG signals.

Building on this dataset, we propose Clinical Anchored Pretraining for PPG (\textbf{CAP}). To bridge the mismatch between pretraining and downstream tasks in terms of signal duration and frequency resolution, CAP introduces Physical-Semantic Unified Patching (PSUP). PSUP performs dynamic resampling and slicing to map heterogeneous long PPG recordings into a sequence of patches with a consistent physical time span, enabling physically meaningful alignment of representation granularity across tasks.

CAP further integrates a set of complementary self-supervised objectives that jointly capture PPG morphology, robustness, and clinical semantics: (1) Morphological Reconstruction, which recovers masked waveform details to preserve fine-grained morphological patterns; (2) Physiological Stability Modeling, which enforces invariance under noise and state perturbations to learn robust features; and (3) Long-term Semantic Anchoring, which introduces a patient-level clinical contrastive loss to align physiological representations with a shared multimodal medical semantic space, thereby mitigating representation drift in long-term monitoring.

Empirically, CAP learns clinically grounded and transferable universal PPG representations, yielding strong performance across multiple downstream monitoring tasks. Our main contributions are summarized as follows:
\begin{itemize}
    \item \textbf{A principled diagnosis of current PPG SSL.} We provide a systematic analysis of why existing self-supervised learning (SSL) paradigms for PPG often fall short of universal transfer, highlighting their reliance on signal-level objectives and the lack of patient-level physiological and clinical context. Motivated by these limitations, we curate a large-scale paired PPG--EHR multimodal pretraining dataset that injects long-horizon patient background information into PPG representation learning.
    
    \item \textbf{CAP: a multimodal pretraining framework for PPG.} Building on this dataset, we propose \textbf{C}linical \textbf{A}nchored \textbf{P}retraining for PPG (CAP). CAP leverages patient-level clinical semantics as a physiological semantic anchor and performs cross-modal contrastive alignment, guiding the PPG encoder beyond waveform fitting toward patient level representations.
    
    \item \textbf{Strong transfer across four heterogeneous downstream tasks.} We conduct comprehensive evaluations on four diverse downstream tasks, including atrial fibrillation detection, heart rate prediction, respiratory rate prediction, and heart rate estimation. CAP consistently outperforms competitive baselines, validating the effectiveness of patient-level clinical anchoring for universal PPG representation learning.
\end{itemize}

\section{Related Works}
\subsection{PPG Self-Supervised Learning}
PPG captures peripheral blood volume pulsations and vascular dynamics, making it highly informative for health monitoring. Accordingly, a substantial literature has studied supervised PPG modeling across diverse downstream tasks such as vital-sign estimation and arrhythmia screening~\cite{ding2024photoplethysmography}. Representative advances include architectures tailored for temporal dynamics (e.g., convolutional recurrent regressors~\cite{ismail2022heart}), training strategies that improve label efficiency and consistency, data augmentation for class imbalance~\cite{ding2023log}, and signal-quality-aware learning to mitigate motion artifacts and acquisition noise (e.g., SQUWA and related variants)~\cite{yan2024squwa, chen2024sparse}. Despite these efforts, PPG learning remains sensitive to noise, motion artifacts, and distribution shifts across devices and cohorts.

To improve generalization and reduce reliance on expensive labels, recent work has moved toward self-supervised and foundation-style pretraining for PPG. These methods typically leverage large-scale recordings and adopt contrastive objectives or waveform-centric reconstruction targets to learn transferable representations that generalize across datasets and tasks~\cite{abbaspourazadlarge,pillaipapagei,saha2025pulse}. In parallel, multimodal and cross-signal pretraining has been explored to transfer representations across PPG, ECG, and other physiological signals; masked-reconstruction paradigms further show promise for multivariate health time-series modeling~\cite{abbaspourazad2024wearable,narayanswamyscaling,yang2023biot}. Collectively, these trends reflect a shift from task-specific pipelines toward more universal representation learning for biosignals.

However, most existing approaches remain largely confined to signal-level objectives and are weakly constrained by real clinical semantics. As a result, the learned representations may capture dataset- or device-specific shortcuts, limiting robustness under clinical distribution shifts. In contrast, our work introduces patient-level clinical context from EHR as semantic anchors for pretraining, complementing signal-only objectives and encouraging representations that better align with clinically meaningful patient states.

\subsection{Representation Learning with Multimodal Medical Data}
Due to the inherent heterogeneity of medical data, multimodal representation learning has become a powerful paradigm for learning semantically grounded features. In radiology, a large body of work has shown that aligning images with free-text reports yields representations that are more robust and clinically meaningful for downstream tasks~\cite{liu2023m,wan2023med,pillaipapagei,ding2024siamquality,chen2025gptppg,huang2025combining}. 

Extending this paradigm to physiological signals is non-trivial~\cite{li2025anchorinv}. Unlike images~\cite{zhou2025mamba}, physiological waveforms exhibit strong temporal structure, complex generative dynamics, and limited human interpretability, which complicates the learning of stable, transferable semantics from local patterns alone~\cite{shen2025physiological}. Nevertheless, recent efforts have begun to pair physiological signals with clinical text, most notably for ECG~\cite{li2025electrocardiogram}. Prior studies demonstrate the potential of jointly modeling ECG with EHR, but many existing designs rely on coarse condition names or weak prompts, which may miss richer clinical attributes and longitudinal context and thus provide limited semantic supervision~\cite{liu2024zero}.

In contrast, multimodal representation learning for PPG remains comparatively under-explored. Most prior work is task-specific and does not provide a general-purpose pretraining framework that supports broad transfer. Our work fills this gap by incorporating patient-level clinical semantics from EHR as supervision for PPG pretraining, enabling reusable universal representations and systematically demonstrating their benefits across diverse downstream tasks.

\section{PPG Multimodal Dataset}
We build our multimodal datasets on \textsc{MC-MED} from emergency-department (ED). ED patients frequently present with undifferentiated complaints and rapidly evolving physiology. Consequently, in addition to EHR, continuous physiological monitoring is crucial for characterizing acute disease trajectories and responses to clinical interventions. To maximize reusability and clinical validity, we curate the paired PPG--text dataset under the guidance of medical experts, following three design principles.

\paragraph{\textbf{Principle 1: Strict Temporal Causality.}}
We enforce strict temporal causality when pairing modalities. For each PPG segment recorded up to time $T_{\mathrm{current}}$, the associated text includes only information available at or before $T_{\mathrm{current}}$ (e.g., prior diagnoses, notes, and orders placed before the segment end time). Any future-dependent information, such as discharge diagnoses, outcomes, or post-hoc summaries determined after the signal recording, is excluded to prevent look-ahead bias.

\paragraph{\textbf{Principle 2: Hierarchical Multimodal Alignment.}}
To reflect the multi-timescale nature of ED visits, we adopt a hierarchical alignment strategy that organizes clinical text into three streams: static context (e.g., demographics), cumulative context (e.g., past medical history and laboratory results within a specified look-back window), and concurrent context (e.g., time-stamped observations and interventions proximal to the PPG segment). This design ensures that textual context evolves alongside the patient’s physiological state captured by PPG.

\paragraph{\textbf{Principle 3: Clinical Quality-Aware Filtering.}}
Given the noisy acquisition conditions in the ED, we apply clinical quality-aware filtering to retain reliable PPG segments. Specifically, we compute signal quality indices (SQIs) for each 300-second segment, and keep only segments that exceed a predefined quality threshold. This filtering reduces the risk of training models on acquisition artifacts and encourages learning physiologically meaningful patterns.

Following these principles, we extract fixed-length PPG segments and their matched clinical entries from \textsc{MC-MED}. Each example is a 5-minute PPG window $x_k$ together with all clinical information available up to the end of that window, denoted as $\mathcal{C}_k$. We then use a pretrained LLM to distill $\mathcal{C}_k$ into a structured EHR document $e_k$:
\begin{equation}
e_k = \mathrm{LLM}(\mathcal{C}_k).
\end{equation}
After SQI-based filtering, we obtain the paired multimodal dataset
\begin{equation}
\mathcal{D} = \{(x_k, e_k)\}_{k=1}^{M},
\end{equation}
where $M$ is the total number of retained PPG--EHR pairs.

\begin{figure*}[h]
  \centering
  \includegraphics[width=\linewidth]{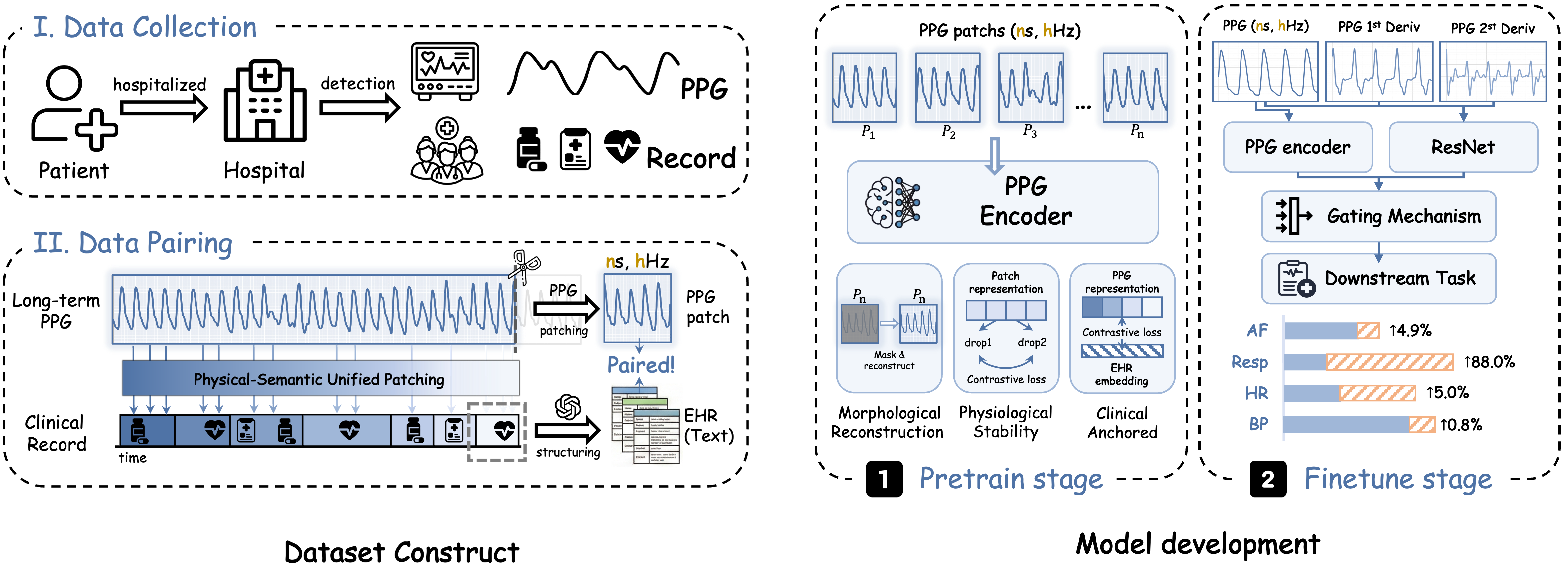}
  \caption{Model architecture overview. In data construct part, (I) hospital-acquired PPG recordings are paired with matched clinical records and (II) converted into downstream-compatible PPG patches via Physical-Semantic Unified Patching. In model development part, (1) Pretraining stage: PPG encoder is trained with three complementary objectives to jointly capture physiological patterns and clinical semantics. (2) Finetune stage: local signal features are extracted from the PPG waveform and its first-derivatives and second-derivatives, while the pretrained PPG encoder provides long-context clinical representations; combining both yields strong performance on downstream tasks.}

  \label{Model Arch}
\end{figure*}

\section{Clinical Anchored Pretraining for PPG}
\subsection{Pretrain Stage}\label{pretrain}
In this section, we present the proposed \textbf{CAP (Clinical Anchored Pretraining)} framework for PPG universal representation learning. The framework is designed to bridge the gap between low-level physiological fluctuations and high-level clinical semantics through a hierarchical alignment process.

\paragraph{Physical-Semantic Unified Patching}
A fundamental challenge in PPG pretraining is the granularity mismatch between long term recordings (e.g., 5-minute pretraining samples) and short term downstream tasks (e.g., 30-second diagnostics). To achieve physical and semantic consistency, we propose the Physical-Semantic Unified Patching mechanism.

Given a target duration $T_d$ and a desired number of sampling points $N_p$ for downstream compatibility, we first define the target sampling frequency as $f_{target} = N_p / T_d$. The raw signal $S_i$ with frequency $f_{raw}$ is resampled to $S'_i$:
\begin{equation}
S'_i = \text{Resample}\left(S_i, f_{raw} \to f_{target}\right)
\end{equation}
The resampled sequence $S'_i \in \mathbb{R}^{C \times L'}$ is then partitioned into $M$ non-overlapping patches $\mathbf{x} = \{x_1, x_2, \dots, x_M\}$, where each patch $x_j \in \mathbb{R}^{C \times N_p}$ corresponds exactly to the physical span required by downstream tasks. This mechanism ensures that the learned representation is invariant to sampling rate variations and temporally aligned with diagnostic windows.

The CAP framework optimizes three cooperative objectives to capture physiological features from micro-morphology to macro-clinical semantics.

\paragraph{\textbf{Morphological Reconstruction Loss}}
To encourage the Patch Encoder $\mathcal{F}_{\theta}$ to learn the fundamental ``grammar'' of PPG waveforms, such as systolic peaks and dicrotic notches, we employ Masked Signal Modeling (MSM). A random binary mask $m \in \{0, 1\}^{N_p}$ with ratio $\rho$ is applied to each patch $x_j$. The masked patch $\tilde{x}_j = x_j \odot (1-m)$ is fed into $\mathcal{F}_{\theta}$ and a lightweight decoder $\mathcal{G}_{\phi}$ to reconstruct the original signal:
\begin{equation}
\hat{x}_j = \mathcal{G}_{\phi}(\mathcal{F}_{\theta}(\tilde{x}_j))
\end{equation}
The morphological reconstruction loss is defined as:
\begin{equation}
\mathcal{L}_{morph} = \mathbb{E}_{x \in \mathbf{x}} \left[ \frac{1}{\|\mathcal{M}\|_0} \sum_{k \in \mathcal{M}} \|x_{j,k} - \hat{x}_{j,k}\|^2 \right]
\end{equation}
where $\mathcal{M}$ is the set of masked indices. 

\paragraph{\textbf{Physiological Stability Loss}}
To ensure that the extracted features are robust against sensor noise and baseline wander, we implement a self-supervised contrastive objective. For a single patch $x_j$, two stochastic latent views $z_j$ and $z'_j$ are generated via independent dropout perturbations within $\mathcal{F}_{\theta}$. The stability loss enforces feature consistency in the latent space:
\begin{equation}
\mathcal{L}_{inv} = -\log \frac{\exp(\text{sim}(z_j, z'_j) / \tau)}{\sum_{k=1}^B \exp(\text{sim}(z_j, z_k) / \tau)}
\end{equation}
where $\text{sim}(\cdot, \cdot)$ denotes cosine similarity and $B$ is the batch size.

\paragraph{\textbf{Clinical Anchored Loss}}
To anchor physiological fluctuations to clinical reality, we align the aggregated global PPG representation $P_i = \text{Agg}(\{z_{j,i}\}_{j=1}^M)$ with the structural EHR embedding $T_i$ derived from the LLM-enhanced text. Using a cross-modal contrastive framework~\cite{radford2021learning}, the loss is defined as:
\begin{equation}
\mathcal{L}_{clin} = \frac{1}{2} \left( \mathcal{L}_{p \to t} + \mathcal{L}_{t \to p} \right)
\end{equation}
By anchoring the signal to long-term clinical outcomes, $\mathcal{L}_{clin}$ mitigates representation shift and imbues the encoder with diagnostic relevance.

\subsection{Finetune Stage}
In the finetuning stage, we adapt CAP to task-specific supervision by combining local waveform dynamics with the long-context clinical semantics learned during pretraining. Given a 5-minute PPG segment $x \in \mathbb{R}^{T}$, we compute its first- and second-order temporal derivatives and construct a three-channel input:
\begin{equation}
\mathbf{s} = \big[x,\ \nabla x,\ \nabla^2 x\big] \in \mathbb{R}^{3 \times T},
\end{equation}
where $\nabla x$ and $\nabla^2 x$ denote the discrete first and second derivatives, respectively. This multi-channel representation explicitly exposes slope and curvature cues, which are often informative for local morphology.

\paragraph{\textbf{Local Morphology Encoder.}}
We feed $\mathbf{s}$ into a ResNet-based local encoder $\mathcal{H}_{\psi}$ to extract task-relevant local features:
\begin{equation}
\ell = \mathcal{H}_{\psi}(\mathbf{s}) \in \mathbb{R}^{d_{\ell}}.
\end{equation}

\paragraph{\textbf{Clinical-Semantic Encoder.}}
In parallel, we reuse the CAP pretrained PPG encoder to obtain a clinically grounded long-context representation. Using the same Physical-Semantic Unified Patching procedure as in pretraining, we map the input segment into $N_p$ patches $\{x_j\}_{j=1}^{N_p}$, and encode each patch with the pretrained patch encoder $\mathcal{F}_{\theta}$:
\begin{equation}
z_j = \mathcal{F}_{\theta}(x_j), \quad j=1,\dots,N_p.
\end{equation}
We then aggregate patch embeddings to obtain a global semantic representation:
\begin{equation}
g = \mathrm{Agg}\big(\{z_j\}_{j=1}^{N_p}\big) \in \mathbb{R}^{d_g}.
\end{equation}
Because $\mathcal{F}_{\theta}$ is pretrained with the clinical anchored loss (Section~\ref{pretrain}), $g$ captures long-term clinical semantics beyond local waveform patterns.

\paragraph{\textbf{Gated Fusion and Prediction Head.}}
To integrate complementary information from local morphology and long-context semantics, we employ a gated fusion module. Specifically, we compute a gate vector $\alpha$ conditioned on both feature types:
\begin{equation}
\alpha = \sigma\!\left(W_{\alpha}\,[\ell; g] + b_{\alpha}\right),
\end{equation}
where $[;]$ denotes concatenation and $\sigma(\cdot)$ is the sigmoid function. The fused representation is computed as
\begin{equation}
h = \alpha \odot g + (1-\alpha)\odot \ell,
\end{equation}
where $\odot$ is the element-wise product. Finally, we apply an activation layer and a task-specific linear head to obtain the prediction:
\begin{equation}
\hat{y} = \mathcal{R}_{\omega}\big(\varphi(h)\big),
\end{equation}
where $\varphi(\cdot)$ denotes a non-linear activation and $\mathcal{R}_{\omega}$ is a prediction head constructed by multilayer perceptron (MLP).

\section{Experiment}
For pretraining, we utilize MedCPT as the text encoder for Electronic Health Records (EHR). During downstream training, we optimize the \texttt{BCEWithLogitsLoss} for AF detection and the Mean Squared Error (MSE) for the three regression tasks (HR/BP/RR). For evaluation, we report the F1-score for AF detection and the Mean Absolute Error (MAE) for all regression tasks.

\noindent\textbf{Experiments settings}
All experiments are implemented in PyTorch~\cite{paszke2019pytorch} and run on 8$\times$ NVIDIA A800 80GB GPUs. For each reported result, we repeat training three times with different random seeds and report 99\% confidence intervals. In CAP pretraining, PPG segments are 5 minutes long at 125\,Hz. In downstream tasks, PPG segments are 30 seconds long at 40\,Hz. Since prior work adopts different evaluation protocols, we explicitly detail the methodology used in each table; CV denotes cross-validation and LoSo denotes leave-one-subject-out validation~\cite{chen2025gptppg}.

\noindent\textbf{Datasets \& Baselines.}
We evaluate CAP on four downstream tasks using widely adopted public benchmarks.

\paragraph{Atrial fibrillation (AF) detection.}
We use the \textsc{Stanford} PPG dataset~\cite{torres2020multi}. We follow the official subject-level train/test split provided by the dataset, ensuring no overlap of individuals between training and testing. We compare against DeepBeat~\cite{torres2020multi}, BayesBeat~\cite{das2022bayesbeat}, ResNeXt~\cite{shen2019ambulatory}, SiamQuality~\cite{ding2024siamquality} and GPT-PPG~\cite{chen2025gptppg}. SiamQuality and GPT-PPG are pretrained (foundation-style) approach trained on large-scale PPG recordings.

\paragraph{Heart rate (HR) estimation.}
We adopt \textsc{WESAD}~\cite{philip2018multimodal} and perform leave-one-subject-out (LOSO) evaluation. Baselines include BeliefPPG~\cite{bieri2023beliefppg}, Deep PPG~\cite{reiss2019deep}, TAPIR~\cite{huang2020robust}, SiamQuality~\cite{ding2024siamquality},PaPaGei~\cite{pillaipapagei} and GPT-PPG~\cite{chen2025gptppg}.

\paragraph{Respiratory rate (RR) estimation.}
We use \textsc{BIDMC}~\cite{pimentel2016toward} and conduct 5-fold cross-validation. For each fold, we enforce subject-disjoint splits between training and validation sets, using an 80\%/20\% ratio. We compare against RRWaveNet~\cite{osathitporn2023rrwavenet}, RespWatch~\cite{dai2021respwatch}, an LSTM baseline~\cite{kumar2022deep}, SiamQuality~\cite{ding2024siamquality} and GPT-PPG~\cite{chen2025gptppg}.

\paragraph{Blood pressure (BP) estimation.}
We evaluate on \textsc{PulseDB}~\cite{wang2023pulsedb}. We create a held-out test set by randomly sampling subjects from the full cohort, and explicitly ensure subject-level disjointness between training and test splits to prevent distribution leakage. We compare against Slapniar~\cite{slapnivcar2019blood}, Kachuee~\cite{kachuee2015cuff}, SiamQuality~\cite{ding2024siamquality} and GPT-PPG~\cite{chen2025gptppg}.

\subsection{AF Detection Results}
As shown in Table~\ref{tab:AF detec}, CAP achieves the best performance on AF detection. Coverage denotes the fraction of the original dataset retained for evaluation, which is typically reduced by quality-based filtering. All results in Table~\ref{tab:AF detec} follow the same official test split; when coverage is below 100\%, evaluation is performed on the filtered subset of the test set rather than the full test set.

\begin{table}[H]
    \centering
    \caption{AF detection results on the \textsc{Stanford}~\cite{torres2020multi}. Coverage denotes the fraction of the original dataset retained for evaluation.}

    \label{tab:AF detec}
    
    \resizebox{0.47\textwidth}{!}{  
        \begin{tabular}{
             l | c | c | c | c
            }
        \specialrule{2.1pt}{0pt}{3pt}
        \multirow{1}{*}{\fontsize{12pt}{14pt}\selectfont Method}
        & \multicolumn{1}{c|}{\fontsize{11pt}{14pt}\selectfont \#Params}
        & \multicolumn{1}{c|}{\fontsize{11pt}{14pt}\selectfont Coverage}
        & \multicolumn{1}{c|}{\fontsize{11pt}{14pt}\selectfont CI (99\%)}
        & \multicolumn{1}{c}{\fontsize{11pt}{14pt}\selectfont F1} \\
    
        \midrule
    
       \multirow{1}{*}{DeepBeat (2020)}
         & 3M 
         & 100\%   
         & \cellcolor[HTML]{E4E5E4}$\pm$0.135
         & \cellcolor[HTML]{EEF3F9}\textbf{0.652} \\

       \multirow{1}{*}{}
         & 3M
         & 27.67\%   
         & \cellcolor[HTML]{E4E5E4}$\pm$0.209
         & \cellcolor[HTML]{F2F6FB}\textbf{0.646}\\

        \specialrule{0.8pt}{0pt}{3pt} 

       \multirow{1}{*}{ResNext (2019)}
         & 5M
         & 100\%  
         & \cellcolor[HTML]{E4E5E4}$\pm$0.118
         & \cellcolor[HTML]{D8E1ED}\textbf{0.684} \\

        \specialrule{0.8pt}{0pt}{3pt} 

       \multirow{1}{*}{BayesBeat (2020)}
         & 2M
         & 100\%  
         & \cellcolor[HTML]{E4E5E4}$\pm$0.099
         & \cellcolor[HTML]{E1E8F2}\textbf{0.671} \\

       \multirow{1}{*}{}
         & 2M
         & 54\%   
         & \cellcolor[HTML]{E4E5E4}$\pm$0.234
         & \cellcolor[HTML]{A8BAD4}\textbf{0.754} \\

        \specialrule{0.8pt}{0pt}{3pt} 

       \multirow{1}{*}{SiamQuality (2024)}
         & 19M
         & 100\%   
         & \cellcolor[HTML]{E4E5E4}$\pm$0.124
         & \cellcolor[HTML]{9DB1CF}\textbf{0.770} \\

        \specialrule{0.8pt}{0pt}{3pt} 

       \multirow{1}{*}{GPT-PPG (2025)}
         & 19M
         & 100\%   
         & \cellcolor[HTML]{E4E5E4}$\pm$0.101
         & \cellcolor[HTML]{C6D3E4}\textbf{0.710} \\
       \multirow{1}{*}{}
         & 85M
         & 100\%   
         & \cellcolor[HTML]{E4E5E4}$\pm$0.095
         & \cellcolor[HTML]{9DB1CF}\textbf{0.770} \\
       \multirow{1}{*}{}
         & 345M
         & 100\%   
         & \cellcolor[HTML]{E4E5E4}$\pm$0.105
         & \cellcolor[HTML]{7894BC}\textbf{0.823} \\
       \multirow{1}{*}{}
         & 1B
         & 100\%   
         & \cellcolor[HTML]{E4E5E4}$\pm$0.116
         & \cellcolor[HTML]{6C8AB5}\textbf{0.841} \\
        \specialrule{0.8pt}{0pt}{3pt} 
        
       \multirow{1}{*}{CAP (Ours)}
         & 130M
         & 100\%   
         & \cellcolor[HTML]{E4E5E4}$\pm$0.010  
         & \cellcolor[HTML]{4F73A6}\textbf{0.883 (Best)} \\    

        \specialrule{1.8pt}{0pt}{3pt} 
        \end{tabular}  
    }
\end{table}

\subsection{Heart Rate Estimation Results}
As shown in Table~\ref{tab:HR estimate}, CAP achieves the best performance on the HR estimation task. We report results on HR estimation task in table 2. TAPIR performs quality-based filtering on the dataset.

\begin{table}[H]
    \centering
    \caption{HR Estimation Results on the \textsc{WESAD}~\cite{philip2018multimodal}.}
    \label{tab:HR estimate}
    
    \resizebox{0.47\textwidth}{!}{  
        \begin{tabular}{
             l | c | c | c | c
            }
        \specialrule{2.1pt}{0pt}{3pt}
        \multirow{1}{*}{\fontsize{12pt}{14pt}\selectfont Method}
        & \multicolumn{1}{c|}{\fontsize{11pt}{14pt}\selectfont \#Params}
        & \multicolumn{1}{c|}{\fontsize{11pt}{14pt}\selectfont Eval}
        & \multicolumn{1}{c|}{\fontsize{11pt}{14pt}\selectfont CI (99\%)}
        & \multicolumn{1}{c}{\fontsize{11pt}{14pt}\selectfont MAE} \\
    
        \midrule
    
       \multirow{1}{*}{BeliefPPG (2023)}
         & 3M 
         & LoSo  
         & \cellcolor[HTML]{E4E5E4}$\pm$2.00
         & \cellcolor[HTML]{5D7EAD}\textbf{4.28} \\

        \specialrule{0.8pt}{0pt}{3pt} 

       \multirow{1}{*}{Deep PPG (2019)}
         & 3M
         & LoSo  
         & \cellcolor[HTML]{E4E5E4}$\pm$3.30
         & \cellcolor[HTML]{F2F6FB}\textbf{7.47}\\

        \specialrule{0.8pt}{0pt}{3pt} 

       \multirow{1}{*}{TAPIR (2020)}
         & 5M
         & LoSo
         & \cellcolor[HTML]{E4E5E4}$\pm$1.40
         & \cellcolor[HTML]{597BAB}\textbf{4.20} \\

        \specialrule{0.8pt}{0pt}{3pt} 

       \multirow{1}{*}{SiamQuality (2024)}
         & 2M
         & LoSo  
         & \cellcolor[HTML]{E4E5E4}$\pm$2.34
         & \cellcolor[HTML]{A8BAD4}\textbf{5.88} \\

        \specialrule{0.8pt}{0pt}{3pt} 

       \multirow{1}{*}{PaPaGei-P (2024)}
         & 2M
         & Random
         & \cellcolor[HTML]{E4E5E4} --
         & \cellcolor[HTML]{FFFFFF} -- \\

        \specialrule{0.8pt}{0pt}{3pt} 

       \multirow{1}{*}{GPT-PPG (2025)}
         & 19M
         & Random 
         & \cellcolor[HTML]{E4E5E4} --
         & \cellcolor[HTML]{FFFFFF} -- \\
       \multirow{1}{*}{}
         & 85M
         & LoSo  
         & \cellcolor[HTML]{E4E5E4} $\pm$2.20
         & \cellcolor[HTML]{92A9C9}\textbf{5.42} \\
       \multirow{1}{*}{}
         & 345M
         & LoSo  
         & \cellcolor[HTML]{E4E5E4}$\pm$2.00
         & \cellcolor[HTML]{8DA5C6}\textbf{5.32} \\
       \multirow{1}{*}{}
         & 1B
         & LoSo  
         & \cellcolor[HTML]{E4E5E4}$\pm$1.80
         & \cellcolor[HTML]{7D98BE}\textbf{4.98} \\
        \specialrule{0.8pt}{0pt}{3pt} 
        
       \multirow{1}{*}{CAP (Ours)}
         & 130M
         & LoSo   
         & \cellcolor[HTML]{E4E5E4}$\pm$1.06
         & \cellcolor[HTML]{4F73A6}\textbf{3.99 (Best)} \\    

        \specialrule{1.8pt}{0pt}{3pt} 
        \end{tabular}  
    }
\end{table}

\subsection{Respiration Rate Estimation Results}
As shown in Table~\ref{tab:Resp estimate}, CAP achieves the best performance on the HR estimation task. Since the dataset does not provide an official fixed test split, we report results using 5-fold CV. CAP consistently outperforms competing methods, even when trained and evaluated without any quality-based filtering.

\begin{table}[H]
    \centering
    \caption{Resp estimation Results on the \textsc{BIDMC}~\cite{pimentel2016toward}.}
    \label{tab:Resp estimate}
    
    \resizebox{0.47\textwidth}{!}{  
        \begin{tabular}{
             l | c | c | c | c
            }
        \specialrule{2.1pt}{0pt}{3pt}
        \multirow{1}{*}{\fontsize{12pt}{14pt}\selectfont Method}
        & \multicolumn{1}{c|}{\fontsize{11pt}{14pt}\selectfont \#Params}
        & \multicolumn{1}{c|}{\fontsize{11pt}{14pt}\selectfont Eval}
        & \multicolumn{1}{c|}{\fontsize{11pt}{14pt}\selectfont CI (99\%)}
        & \multicolumn{1}{c}{\fontsize{11pt}{14pt}\selectfont MAE} \\
    
        \midrule
    
       \multirow{1}{*}{RRWaveNet (2023)}
         & 3M 
         & LOOCV
         & \cellcolor[HTML]{E4E5E4}$\pm$1.01
         & \cellcolor[HTML]{C7D4E5}\textbf{1.66} \\

        \specialrule{0.8pt}{0pt}{3pt} 

       \multirow{1}{*}{RespWatch (2021)}
         & 3M
         & 5-fold CV
         & \cellcolor[HTML]{E4E5E4}$\pm$2.80
         & \cellcolor[HTML]{C7D4E5}\textbf{1.66}\\

        \specialrule{0.8pt}{0pt}{3pt} 

       \multirow{1}{*}{LSTM (2022)}
         & 5M
         & Random
         & \cellcolor[HTML]{E4E5E4}$\pm$2.09
         & \cellcolor[HTML]{BCCADF}\textbf{1.51} \\

        \specialrule{0.8pt}{0pt}{3pt} 

       \multirow{1}{*}{SiamQuality (2024)}
         & 2M
         & 5-fold CV
         & \cellcolor[HTML]{E4E5E4}$\pm$0.01
         & \cellcolor[HTML]{8CA4C6}\textbf{0.89} \\

        \specialrule{0.8pt}{0pt}{3pt} 

       \multirow{1}{*}{GPT-PPG (2025)}
         & 19M
         & 5-fold CV
         & \cellcolor[HTML]{E4E5E4}$\pm$0.69
         & \cellcolor[HTML]{F2F6FB}\textbf{2.21} \\
       \multirow{1}{*}{}
         & 85M
         & 5-fold CV
         & \cellcolor[HTML]{E4E5E4}$\pm$0.56
         & \cellcolor[HTML]{C6D2E4}\textbf{1.64} \\
       \multirow{1}{*}{}
         & 345M
         & 5-fold CV
         & \cellcolor[HTML]{E4E5E4}$\pm$0.48
         & \cellcolor[HTML]{AEBFD7}\textbf{1.33} \\
       \multirow{1}{*}{}
         & 1B
         & 5-fold CV
         & \cellcolor[HTML]{E4E5E4}$\pm$0.21
         & \cellcolor[HTML]{8FA6C7}\textbf{0.93} \\
        \specialrule{0.8pt}{0pt}{3pt} 
        
       \multirow{1}{*}{CAP (Ours)}
         & 130M
         & 5-fold CV
         & \cellcolor[HTML]{E4E5E4}$\pm$0.05
         & \cellcolor[HTML]{4F73A6}\textbf{0.11 (Best)} \\    

        \specialrule{1.8pt}{0pt}{3pt} 
        \end{tabular}  
    }
\end{table}

\subsection{Blood Pressure Estimation Results}
As shown in Table~\ref{tab:Resp estimate}, CAP achieves the best performance on the HR estimation task. We evaluate on \textsc{PulseDB}~\cite{wang2023pulsedb}, a curated dataset derived from \textsc{MIMIC-III}~\cite{johnson2016mimic} and \textsc{VitalDB}~\cite{lee2022vitaldb}. We compare against prior methods reported on \textsc{MIMIC-III}; however, since different works adopt non-matching signal filtering and preprocessing pipelines, these comparisons should be interpreted as a reference rather than a strictly controlled benchmark. Given the large scale of \textsc{PulseDB} (483{,}844 samples), we use a single held-out test split instead of 5-fold cross-validation, and keep the randomly sampled test set fixed across all our runs for consistency.

\begin{table}[H]
    \centering
    \caption{BP Estimation Results on the \textsc{PulseDB}~\cite{wang2023pulsedb}.}
    \label{tab:BP estimate}
    
    \resizebox{0.47\textwidth}{!}{  
        \begin{tabular}{
             l | c | c | c | c
            }
        \specialrule{2.1pt}{0pt}{3pt}
        \multirow{1}{*}{\fontsize{12pt}{14pt}\selectfont Method}
        & \multicolumn{1}{c|}{\fontsize{11pt}{14pt}\selectfont \#Params}
        & \multicolumn{1}{c|}{\fontsize{11pt}{14pt}\selectfont Eval}
        & \multicolumn{1}{c|}{\fontsize{11pt}{14pt}\selectfont CI (99\%)}
        & \multicolumn{1}{c}{\fontsize{11pt}{14pt}\selectfont MAE} \\
    
        \midrule
    
       \multirow{1}{*}{Slapniar (2019)}
         & 3M 
         & LoSo
         & \cellcolor[HTML]{E4E5E4}$\pm$6.35
         & \cellcolor[HTML]{839DC1}\textbf{9.43} \\

        \specialrule{0.8pt}{0pt}{3pt} 

       \multirow{1}{*}{Kachuee  (2015)}
         & 3M
         & Random
         & \cellcolor[HTML]{E4E5E4}$\pm$8.92
         & \cellcolor[HTML]{F2F6FB}\textbf{12.38}\\

        \specialrule{0.8pt}{0pt}{3pt} 

       \multirow{1}{*}{SiamQuality (2024)}
         & 2M
         & 5-fold CV
         & \cellcolor[HTML]{E4E5E4}$\pm$6.93
         & \cellcolor[HTML]{6484B1}\textbf{8.60} \\

        \specialrule{0.8pt}{0pt}{3pt} 

       \multirow{1}{*}{GPT-PPG (2025)}
         & 19M
         & Random
         & \cellcolor[HTML]{E4E5E4}$\pm$5.98
         & \cellcolor[HTML]{7894BB}\textbf{9.13} \\
       \multirow{1}{*}{}
         & 85M
         & Random
         & \cellcolor[HTML]{E4E5E4}$\pm$7.56
         & \cellcolor[HTML]{6C8AB5}\textbf{8.81} \\
       \multirow{1}{*}{}
         & 345M
         & Random
         & \cellcolor[HTML]{E4E5E4}$\pm$6.80
         & \cellcolor[HTML]{5D7EAD}\textbf{8.42} \\
       \multirow{1}{*}{}
         & 1B
         & Random
         & \cellcolor[HTML]{E4E5E4}$\pm$7.09
         & \cellcolor[HTML]{5275A8}\textbf{8.12} \\
        \specialrule{0.8pt}{0pt}{3pt} 
        
       \multirow{1}{*}{CAP (Ours)}
         & 130M
         & Random
         & \cellcolor[HTML]{E4E5E4}$\pm$6.39
         & \cellcolor[HTML]{4F73A6}\textbf{8.04 (Best)} \\    

        \specialrule{1.8pt}{0pt}{3pt} 
        \end{tabular}  
    }
\end{table}

\section{Ablation Study}
In this section, we perform comprehensive ablation studies on the key components choices, and report the average performance.

\noindent\textbf{Pretrain Stage Loss.}
In the pretraining stage, CAP optimizes three objectives to learn PPG representations. As shown in Table~\ref{tab:Ablating Loss Function}, we conduct a detailed ablation study to quantify the contribution of each loss term. We first evaluate each objective in isolation. The results show that the clinical anchored loss $\mathcal{L}_{clin}$, which injects long-horizon patient context, provides the largest performance gain, directly supporting our motivation. We then progressively remove each loss component from the full model; performance consistently degrades in all cases, indicating that each term contributes to the overall effectiveness of CAP.

\begin{table}[H]
    \centering
    \caption{Ablating Loss Function.}
    \label{tab:Ablating Loss Function}
    \resizebox{0.42\textwidth}{!}{  
        \begin{tabular}{
             c  c  c | c  c
            }
        \specialrule{2.1pt}{0pt}{3pt}
        \multicolumn{1}{c}{\fontsize{10pt}{14pt}\selectfont $\mathcal{L}_{morph}$}
        & \multicolumn{1}{c}{\fontsize{10pt}{14pt}\selectfont $\mathcal{L}_{inv}$}
        & \multicolumn{1}{c|}{\fontsize{10pt}{14pt}\selectfont $\mathcal{L}_{clin}$}
        & \multicolumn{1}{c}{\fontsize{10pt}{14pt}\selectfont AF Detection}
        & \multicolumn{1}{c}{\fontsize{10pt}{14pt}\selectfont Resp Estimation} \\
    
        \midrule
    
           $\checkmark$
         & 
         & 
         & 0.72 \cibadge{\scriptsize\textcolor{cifg}{$\pm$0.03}}  
         & 0.76 \cibadge{\scriptsize\textcolor{cifg}{$\pm$0.22}}   \\

         & $\checkmark$
         & 
         & 0.69 \cibadge{\scriptsize\textcolor{cifg}{$\pm$0.05}}  
         & 1.23 \cibadge{\scriptsize\textcolor{cifg}{$\pm$1.05}}   \\

         & 
         & $\checkmark$
         & 0.72 \cibadge{\scriptsize\textcolor{cifg}{$\pm$0.06}}  
         & 1.66 \cibadge{\scriptsize\textcolor{cifg}{$\pm$0.31}}   \\

        \specialrule{0.8pt}{0pt}{3pt} 

           $\checkmark$
         & 
         & $\checkmark$
         & 0.75 \cibadge{\scriptsize\textcolor{cifg}{$\pm$0.02}}  
         & 0.36 \cibadge{\scriptsize\textcolor{cifg}{$\pm$0.23}}   \\

         & $\checkmark$
         & $\checkmark$
         & 0.72 \cibadge{\scriptsize\textcolor{cifg}{$\pm$0.00}}  
         & 1.28 \cibadge{\scriptsize\textcolor{cifg}{$\pm$0.41}}   \\

           $\checkmark$
         & $\checkmark$
         & 
         & 0.79 \cibadge{\scriptsize\textcolor{cifg}{$\pm$0.04}}  
         & 0.75 \cibadge{\scriptsize\textcolor{cifg}{$\pm$0.45}}   \\

        \specialrule{0.8pt}{0pt}{0pt} 
        
           \cellcolor[HTML]{E4E5E4}$\checkmark$
         & \cellcolor[HTML]{E4E5E4}$\checkmark$
         & \cellcolor[HTML]{E4E5E4}$\checkmark$
         & \cellcolor[HTML]{E4E5E4}0.88 \cibadge{\scriptsize\textcolor{cifg}{$\pm$0.01}}  
         & \cellcolor[HTML]{E4E5E4}0.11 \cibadge{\scriptsize\textcolor{cifg}{$\pm$0.05}}   \\

        \specialrule{1.8pt}{0pt}{3pt} 
        \end{tabular}  
    }
\end{table}

\noindent\textbf{Text Encoder for $\mathcal{L}_{clin}$.}
Computing the clinical anchored loss $\mathcal{L}_{clin}$ requires embedding the paired EHR text with a pretrained language model. Table~\ref{tab:Text Encoder} summarizes the impact of different text encoders; among all candidates, MedCPT~\cite{jin2023medcpt} achieves the best performance on both tasks.

\begin{table}[H]
    \centering
    \caption{Effects of Text Encoder.}
    \label{tab:Text Encoder}
    \resizebox{0.40\textwidth}{!}{  
        \begin{tabular}{
             c | c  c
            }
        \specialrule{2.1pt}{0pt}{3pt}
        \multicolumn{1}{c}{\fontsize{10pt}{14pt}\selectfont EHR Encoder}
        & \multicolumn{1}{c}{\fontsize{10pt}{14pt}\selectfont AF Detection}
        & \multicolumn{1}{c}{\fontsize{10pt}{14pt}\selectfont Resp Estimation} \\
    
        \midrule
    
         BioClinicalBERT~\cite{alsentzer2019publicly}
         & 0.84 \cibadge{\scriptsize\textcolor{cifg}{$\pm$0.03}}  
         & 0.22 \cibadge{\scriptsize\textcolor{cifg}{$\pm$0.10}}   \\

         PubMedBERT~\cite{gu2021domain}
         & 0.81 \cibadge{\scriptsize\textcolor{cifg}{$\pm$0.05}}  
         & 0.14 \cibadge{\scriptsize\textcolor{cifg}{$\pm$0.03}}   \\

        \specialrule{0.8pt}{0pt}{0pt} 

         \cellcolor[HTML]{E4E5E4}Med-CPT (Ours)~\cite{jin2023medcpt}
         & \cellcolor[HTML]{E4E5E4}0.88 \cibadge{\scriptsize\textcolor{cifg}{$\pm$0.01}}  
         & \cellcolor[HTML]{E4E5E4}0.11 \cibadge{\scriptsize\textcolor{cifg}{$\pm$0.05}}   \\

        \specialrule{1.8pt}{0pt}{3pt} 
        \end{tabular}  
    }
\end{table}

\noindent\textbf{Data Augmentation Strategies for $\mathcal{L}_{inv}$.}
Computing $\mathcal{L}_{inv}$ requires constructing positive/negative pairs via data augmentation. We evaluate several pairing strategies in Table~\ref{tab:Data Augmentation Strategies}. In particular, applying dropout in the representation space is less disruptive than augmenting the raw waveform, and thus better preserves physiologically meaningful information in the signal.

\begin{table}[H]
    \centering
    \caption{Data Augmentation Strategies.}
    \label{tab:Data Augmentation Strategies}
    \resizebox{0.40\textwidth}{!}{  
        \begin{tabular}{
             c | c  c
            }
        \specialrule{2.1pt}{0pt}{3pt}
        \multicolumn{1}{c}{\fontsize{10pt}{14pt}\selectfont EHR Encoder}
        & \multicolumn{1}{c}{\fontsize{10pt}{14pt}\selectfont AF Detection}
        & \multicolumn{1}{c}{\fontsize{10pt}{14pt}\selectfont Resp Estimation} \\
    
        \midrule
    
         Cutout
         & 0.72 \cibadge{\scriptsize\textcolor{cifg}{$\pm$0.12}}  
         & 1.48 \cibadge{\scriptsize\textcolor{cifg}{$\pm$0.60}}   \\

         Drop
         & 0.82 \cibadge{\scriptsize\textcolor{cifg}{$\pm$0.05}}  
         & 1.07 \cibadge{\scriptsize\textcolor{cifg}{$\pm$0.21}}   \\

         Gaussian noise
         & 0.77 \cibadge{\scriptsize\textcolor{cifg}{$\pm$0.02}}  
         & 0.92 \cibadge{\scriptsize\textcolor{cifg}{$\pm$0.42}}   \\

        \specialrule{0.8pt}{0pt}{0pt} 

         \cellcolor[HTML]{E4E5E4}Latent Dropout (Ours)
         & \cellcolor[HTML]{E4E5E4}0.88 \cibadge{\scriptsize\textcolor{cifg}{$\pm$0.01}}  
         & \cellcolor[HTML]{E4E5E4}0.11 \cibadge{\scriptsize\textcolor{cifg}{$\pm$0.05}}   \\

        \specialrule{1.8pt}{0pt}{3pt} 
        \end{tabular}  
    }
\end{table}

\noindent\textbf{Dropout Ratio}
In the previous section, we verified the effectiveness of dropout-based view generation. Here, we further study the impact of the dropout ratio. As shown in Table~\ref{tab:Dropout Ratio}, CAP is relatively sensitive to this hyperparameter and achieves the best overall performance at a dropout ratio of 0.2; we therefore use 0.2 in all experiments.

\begin{table}[H]
    \centering
    \caption{Dropout Ratio.}
    \label{tab:Dropout Ratio}
    \resizebox{0.35\textwidth}{!}{  
        \begin{tabular}{
             c | c  c
            }
        \specialrule{2.1pt}{0pt}{3pt}
        \multicolumn{1}{c}{\fontsize{10pt}{14pt}\selectfont Drop Ratio}
        & \multicolumn{1}{c}{\fontsize{10pt}{14pt}\selectfont AF Detection}
        & \multicolumn{1}{c}{\fontsize{10pt}{14pt}\selectfont Resp Estimation} \\
    
        \midrule
    
         0.1
         & 0.85 \cibadge{\scriptsize\textcolor{cifg}{$\pm$0.02}}  
         & 0.14 \cibadge{\scriptsize\textcolor{cifg}{$\pm$0.02}}   \\

         0.5
         & 0.82 \cibadge{\scriptsize\textcolor{cifg}{$\pm$0.03}}  
         & 0.15 \cibadge{\scriptsize\textcolor{cifg}{$\pm$0.02}}   \\

        \specialrule{0.8pt}{0pt}{0pt} 
         \cellcolor[HTML]{E4E5E4}0.2
         & \cellcolor[HTML]{E4E5E4}0.88 \cibadge{\scriptsize\textcolor{cifg}{$\pm$0.01}}  
         & \cellcolor[HTML]{E4E5E4}0.11 \cibadge{\scriptsize\textcolor{cifg}{$\pm$0.05}}   \\

        \specialrule{1.8pt}{0pt}{3pt} 
        \end{tabular}  
    }
\end{table}

\noindent\textbf{Signal branch in the finetuning stage.}
In finetuning, we augment the raw PPG waveform with its first- and second-order derivatives to expand the local feature space. This design is clinically motivated, as derivative signals capture physiologically meaningful dynamics (e.g., slope and curvature) and provide complementary local cues that can help adaptation across tasks and domains. In this ablation, we aim to answer two questions: (1) whether the clinical-semantic information provided by the pretrained PPG encoder contributes beyond a purely task-specific ResNet backbone; and (2) whether each signal component (raw PPG, first derivative, second derivative) is necessary rather than redundant.

\textbf{Q1: Does the pretrained PPG embedding provide non-trivial gains?}
Using only the pretrained PPG embedding already achieves competitive performance (Row~1), but it is limited by insufficient sensitivity to task-specific local patterns. In contrast, using only the ResNet-based local features (Rows~2--5) does not surpass the full CAP model. When we combine the ResNet branch with the pretrained PPG embedding (Rows~6--8), performance consistently improves across settings. These results indicate that the pretrained embedding contributes complementary long-context clinical semantics and yields measurable gains beyond local feature learning.

\textbf{Q2: Are the derivative channels useful?}
Comparing Rows~2--9, removing any single signal component leads to a performance drop, demonstrating that both the first- and second-order derivatives provide meaningful information. This supports our design choice of incorporating derivative signals as informative local features in the finetuning stage.

\begin{table}[H]
    \centering
    \caption{Ablating Signal branch.}
    \label{tab:Ablating Signal branch}
    \resizebox{0.48\textwidth}{!}{  
        \begin{tabular}{
             c c  c  c | c  c
            }
        \specialrule{2.1pt}{0pt}{3pt}
        \multicolumn{1}{c}{\fontsize{10pt}{14pt}\selectfont PPG emb}
        & \multicolumn{1}{c}{\fontsize{10pt}{14pt}\selectfont PPG}
        & \multicolumn{1}{c}{\fontsize{10pt}{14pt}\selectfont $1^{st} dev$}
        & \multicolumn{1}{c|}{\fontsize{10pt}{14pt}\selectfont $2^{st} dev$}
        & \multicolumn{1}{c}{\fontsize{10pt}{14pt}\selectfont AF Detection}
        & \multicolumn{1}{c}{\fontsize{10pt}{14pt}\selectfont Resp Estimation} \\
    
        \midrule

           $\checkmark$
         &
         & 
         & 
         & 0.68 \cibadge{\scriptsize\textcolor{cifg}{$\pm$0.02}}  
         & 0.23 \cibadge{\scriptsize\textcolor{cifg}{$\pm$0.01}}   \\

        \specialrule{0.8pt}{0pt}{3pt}

         & $\checkmark$
         & 
         & 
         & 0.79 \cibadge{\scriptsize\textcolor{cifg}{$\pm$0.02}}  
         & 0.18 \cibadge{\scriptsize\textcolor{cifg}{$\pm$0.01}}   \\

         & 
         & $\checkmark$
         & 
         & 0.73 \cibadge{\scriptsize\textcolor{cifg}{$\pm$0.05}}  
         & 0.20 \cibadge{\scriptsize\textcolor{cifg}{$\pm$0.02}}   \\

         & 
         & 
         & $\checkmark$
         & 0.74 \cibadge{\scriptsize\textcolor{cifg}{$\pm$0.00}}  
         & 0.19 \cibadge{\scriptsize\textcolor{cifg}{$\pm$0.01}}   \\

         & $\checkmark$
         & $\checkmark$
         & $\checkmark$
         & 0.80 \cibadge{\scriptsize\textcolor{cifg}{$\pm$0.06}}  
         & 0.18 \cibadge{\scriptsize\textcolor{cifg}{$\pm$0.05}}   \\

        \specialrule{0.8pt}{0pt}{3pt} 

           $\checkmark$
         & $\checkmark$
         & 
         & 
         & 0.84 \cibadge{\scriptsize\textcolor{cifg}{$\pm$0.01}}  
         & 0.14 \cibadge{\scriptsize\textcolor{cifg}{$\pm$0.01}}   \\

           $\checkmark$
         & 
         & $\checkmark$
         & 
         & 0.79 \cibadge{\scriptsize\textcolor{cifg}{$\pm$0.03}}  
         & 0.18 \cibadge{\scriptsize\textcolor{cifg}{$\pm$0.00}}   \\

           $\checkmark$
         & 
         & 
         & $\checkmark$
         & 0.82 \cibadge{\scriptsize\textcolor{cifg}{$\pm$0.01}}  
         & 0.13 \cibadge{\scriptsize\textcolor{cifg}{$\pm$0.00}}   \\

        \specialrule{0.8pt}{0pt}{0pt} 
        
           \cellcolor[HTML]{E4E5E4}$\checkmark$
         & \cellcolor[HTML]{E4E5E4}$\checkmark$
         & \cellcolor[HTML]{E4E5E4}$\checkmark$
         & \cellcolor[HTML]{E4E5E4}$\checkmark$
         & \cellcolor[HTML]{E4E5E4}0.88 \cibadge{\scriptsize\textcolor{cifg}{$\pm$0.01}}   
         & \cellcolor[HTML]{E4E5E4}0.11 \cibadge{\scriptsize\textcolor{cifg}{$\pm$0.05}}   \\

        \specialrule{1.8pt}{0pt}{3pt} 
        \end{tabular}  
    }
\end{table}

\noindent\textbf{Feature Extractor in Finetune stage}
Table~\ref{tab:Feature Extractor} outlines the ablation study for two PPG feature extractor networks (i.e., backbones) in finetune stage: the CNN-based ResNet18~\cite{he2016deep} and the transformer-based ViT-Tiny~\cite{dosovitskiy2020image}.

\begin{table}[H]
    \centering
    \caption{Feature Extractor.}
    \label{tab:Feature Extractor}
    \resizebox{0.40\textwidth}{!}{  
        \begin{tabular}{
             c | c  c
            }
        \specialrule{2.1pt}{0pt}{3pt}
        \multicolumn{1}{c}{\fontsize{10pt}{14pt}\selectfont Feature Extractor}
        & \multicolumn{1}{c}{\fontsize{10pt}{14pt}\selectfont AF Detection}
        & \multicolumn{1}{c}{\fontsize{10pt}{14pt}\selectfont Resp Estimation} \\
    
        \midrule
           
         ViT
         & 0.84 \cibadge{\scriptsize\textcolor{cifg}{$\pm$0.04}}  
         & 0.12 \cibadge{\scriptsize\textcolor{cifg}{$\pm$0.02}}   \\

        \specialrule{0.8pt}{0pt}{0pt} 

         \cellcolor[HTML]{E4E5E4}ResNet (Ours)
         & \cellcolor[HTML]{E4E5E4}0.88 \cibadge{\scriptsize\textcolor{cifg}{$\pm$0.01}}  
         & \cellcolor[HTML]{E4E5E4}0.11 \cibadge{\scriptsize\textcolor{cifg}{$\pm$0.05}}   \\

        \specialrule{1.8pt}{0pt}{3pt} 
        \end{tabular}  
    }
\end{table}

\noindent\textbf{Gating Mechanism.}
To further elucidate the complementary dynamics between the pre-trained global semantics ($g$) and the local morphological features ($\ell$), we conduct an in-depth empirical investigation into the learned gating vector $\alpha$. Recall that the fused dual-path representation $h$ is formulated as:
\begin{equation}
    h = \alpha \cdot g + (1 - \alpha) \cdot \ell,
\end{equation}
where $\alpha \in [0, 1]$ adaptively balances the contribution of each modality. A higher value of $\alpha$ indicates a heavier reliance on pre-trained global EHR-aligned semantics, whereas a lower value prioritizes fine-grained local PPG morphology.

We compute the empirical mean of $\alpha$ across the test sets of the four downstream tasks to observe how the model modulates its focus based on clinical objectives. As summarized in Table~\ref{tab:task_alpha}, the model exhibits distinct task-driven adaptation. For instance, Atrial Fibrillation (AF) detection yields the highest mean $\alpha$ ($0.55$), indicating that the identification of arrhythmic episodes inherently demands broader contextual and long-term semantic cues. Conversely, tasks like Respiratory Rate (Resp) estimation exhibit a lower mean $\alpha$ ($0.42$), reflecting a stronger dependency on localized, cycle-to-cycle waveform variations.

\begin{table}[h]
\centering
\caption{Mean Gating Vector $\alpha$ Across Different Downstream Tasks}
\label{tab:task_alpha}
\begin{tabular}{lc}
\toprule
\textbf{Task} & \textbf{Mean $\alpha$} \\
\midrule
Blood Pressure       & $0.47$ \\
Atrial Fibrillation  & $0.55$ \\
Heart Rate       & $0.49$ \\
Respiratory Rate  & $0.42$ \\
\bottomrule
\end{tabular}
\end{table}

To evaluate the gating mechanism's resilience against domain noise, we utilize the \texttt{pyPPG} toolbox to stratify the AF test samples into four distinct groups based on their Signal Quality Index (SQI). 

Table~\ref{tab:sqi_alpha} presents the sample distributions and the corresponding mean $\alpha$ for each quality range. A clear monotonic trend is observed: as the signal quality degrades from the pristine range ($90\text{--}100\%$) to the severely corrupted range ($<50\%$), the mean $\alpha$ significantly increases from $0.43$ to $0.61$. This empirical behavior demonstrates that when local morphological features become unreliable due to motion artifacts or low perfusion, the model automatically increases its reliance on the robust, pre-trained long-context semantics to compensate for the information loss. Conversely, for high-quality signals, the model naturally prioritizes fine-grained local features. This dynamic adaptation validates the necessity and the intrinsic complementary nature of our dual-path architectural design.

\begin{table}[h]
\centering
\caption{Gating Behavior Tracking Across Different Signal Quality Index (SQI) Ranges for AF Detection}
\label{tab:sqi_alpha}
\begin{tabular}{lcc}
\toprule
\textbf{Quality Range (SQI)} & \textbf{Sample Count} & \textbf{Mean $\alpha$} \\
\midrule
$90\text{--}100\%$ & 16,226 & $0.43$ \\
$70\text{--}90\%$  & 10,729 & $0.51$ \\
$50\text{--}70\%$  & 7,218  & $0.53$ \\
$<50\%$            & 3,974  & $0.61$ \\
\bottomrule
\end{tabular}
\end{table}

\begin{figure}[h]
  \centering
  \includegraphics[width=\linewidth]{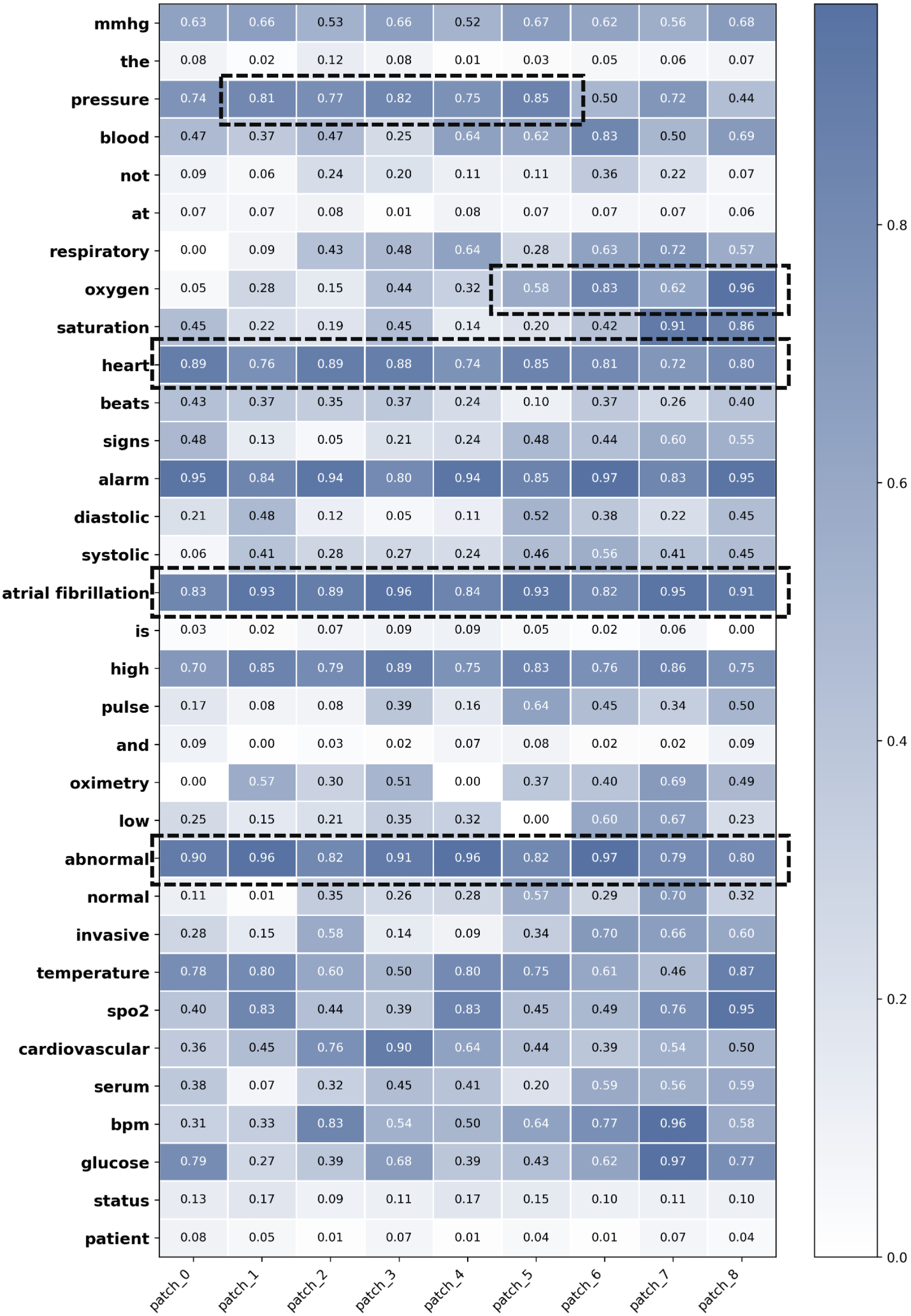}
  \caption{Text embedding visualization on AF task.}
  \label{Text embedding visualization}
\end{figure}

\section{Analysis}
In this section, we aim to improve the interpretability of CAP through complementary analyses from multiple perspectives. Specifically, we examine (i) text--signal interaction attention, (ii) signal-level attention over the PPG waveform, and (iii) UMAP visualizations of the learned representation space. Together, these analyses provide insight into how CAP leverages clinical anchoring and help elucidate the mechanisms behind its empirical gains.

\subsection{Text embedding visualization}
In this section, we provide an interpretability analysis of the learned cross-modal alignment. Specifically, we freeze the pretrained weights of the PPG encoder and MedCPT, and extract embeddings for paired PPG signals and EHR reports from the training set. This yields patch-level representations for PPG and token-level representations for text. We then compute a cross-modal attention map based on the similarity between text tokens and PPG patches, which characterizes how each token aligns with different temporal segments of the PPG waveform. Let $\mathbf{t}_i$ denote the $\ell_2$-normalized embedding of the $i$-th text token and $\mathbf{p}_j$ denote the $\ell_2$-normalized embedding of the $j$-th PPG patch. The token-to-patch cross-attention is defined as:
\begin{equation}
a_{ij}=\frac{\exp\left(\mathbf{t}_i^\top\mathbf{p}_j/\tau\right)}{\sum_{k=1}^{P}\exp\left(\mathbf{t}_i^\top\mathbf{p}_k/\tau\right)},
\end{equation}
where $\tau$ is a temperature hyperparameter and $P$ is the number of valid PPG patches. We visualize the resulting attention matrix $\mathbf{A}=[a_{ij}]$ in Figure~\ref{Text embedding visualization}. As shown, the model assigns high attention weights to clinically meaningful keywords in the EHR text, which are consistent with MedCPT's medical semantics. This observation suggests that the pretraining objective successfully injects clinical information into the PPG representations, facilitating cross-modal semantic matching between physiological signals and textual descriptions.

\begin{figure*}[h]
  \centering
  \includegraphics[width=\linewidth]{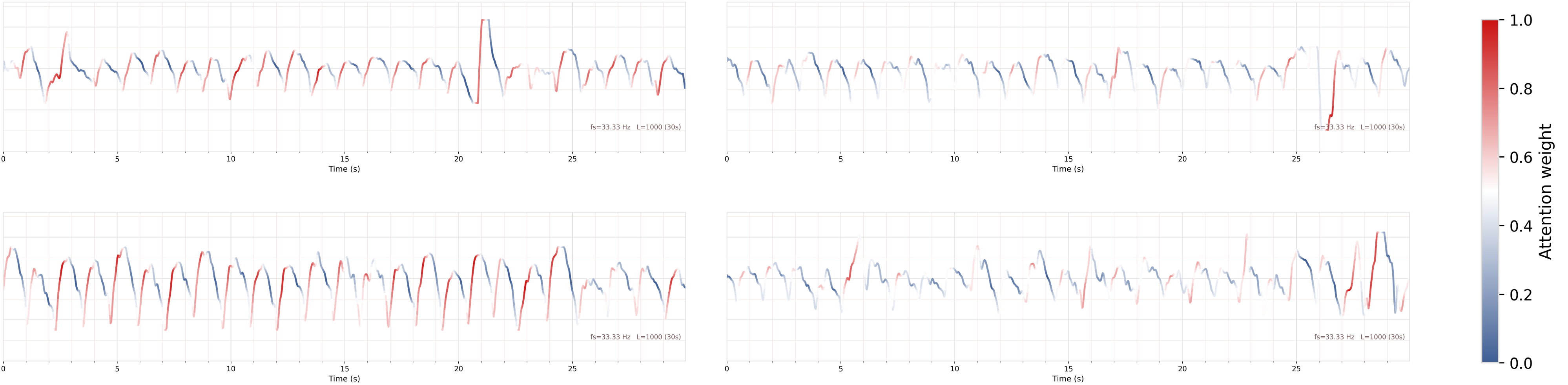}
  \caption{PPG signal attention visualization.}
  \label{signal attn visual}
\end{figure*}

\begin{figure}[t]
    \centering
    \begin{subfigure}[t]{0.49\columnwidth}
        \centering
        \includegraphics[width=\linewidth]{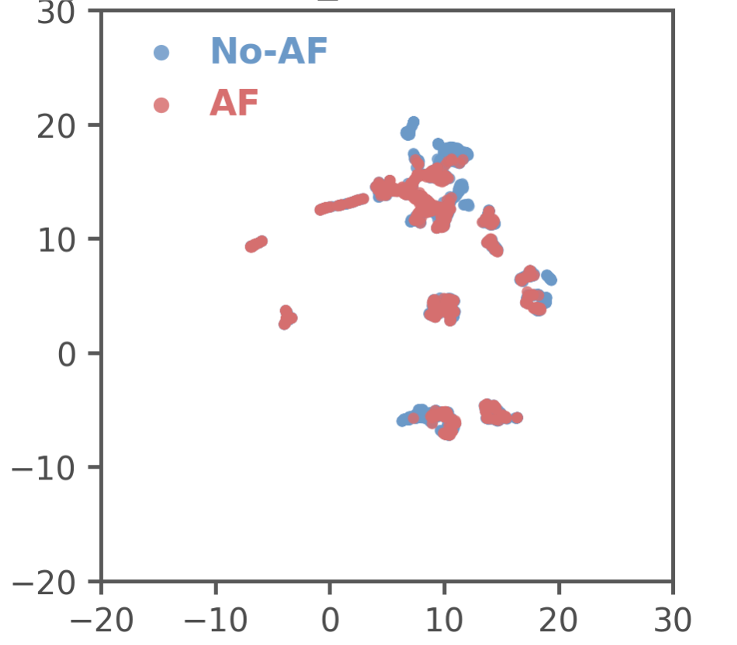}
        \caption{no EHR}
    \end{subfigure}\hfill
    \begin{subfigure}[t]{0.49\columnwidth}
        \centering
        \includegraphics[width=\linewidth]{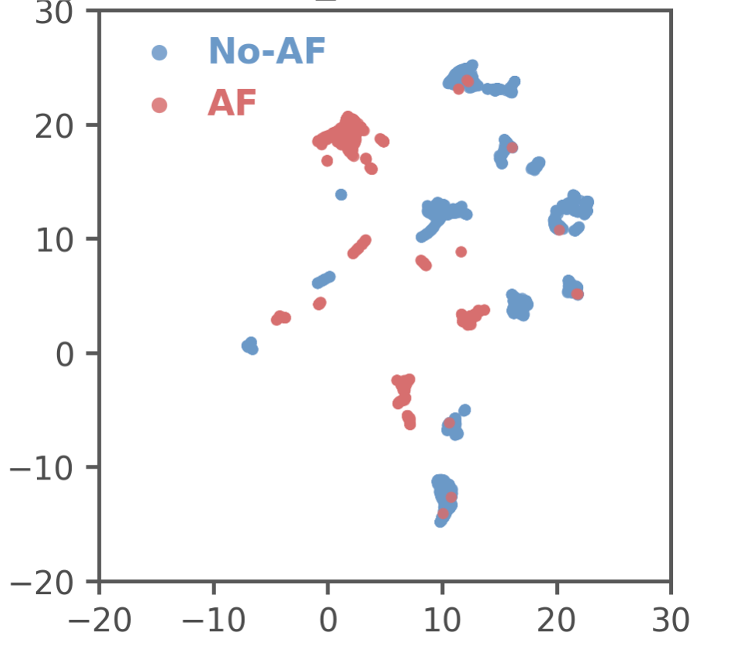}
        \caption{with EHR}
    \end{subfigure}
    \caption{\textbf{EHR supervised t-SNE Visualization on AF task.}}
    \label{UMAP_AF}
\end{figure}

\subsection{PPG Signal Attention Visualization}
To provide an intuitive view of how the model interprets PPG in downstream tasks, we visualize attention over the input signal. Specifically, we use the PPG encoder to obtain patch-level representation weights, and then map these weights back to the corresponding time indices to construct a time-resolved attention profile over the waveform. We select several low-noise examples for clearer visualization.

As shown in Figure~\ref{signal attn visual}, the model consistently assigns higher attention to the rising phase of the PPG waveform and the peak-to-peak region. This observation is physiologically plausible: the rising waveform is closely related to the arterial pulse arrival and is sensitive to vascular compliance and peripheral resistance, while peak-to-peak timing reflects cardiac cycle dynamics and is directly associated with heart rate variability. The attention patterns therefore suggest that CAP tends to focus on clinically meaningful components of PPG rather than noise-dominated segments, supporting the interpretability and medical relevance of the learned representations.

\subsection{UMAP analysis}
We use UMAP to visualize the representation space learned by the PPG encoder, comparing encoders trained with unimodal pretraining versus our multimodal (EHR-anchored) pretraining. We take AF detection as an illustrative example, since class separation is more directly observable in a classification setting. As shown in Figure~\ref{UMAP_AF}, the multimodally trained encoder exhibits substantially clearer separation between AF-positive and AF-negative samples. This improved clustering structure indicates that the EHR-derived supervision provides clinically meaningful constraints and helps shape a more discriminative representation space.

\section{Conclusion}
In this work, we argue that PPG representation learning should be patient-centric, moving beyond short-window, signal-level pattern matching toward modeling patient-level physiological states. We identify three key limitations of current PPG self-supervised learning, including geometric-similarity bias, augmentation-induced semantic corruption, and limited temporal-scale awareness. To address these issues, we introduce patient-level clinical history as semantic supervision.

To support this paradigm, we curate a paired PPG--EHR multimodal dataset from \textbf{2,279} patients, where EHR is distilled from heterogeneous clinical records by a pretrained language model. Building on this dataset, we propose CAP, a clinical anchored pretraining framework for PPG. CAP combines Physical-Semantic Unified Patching to reconcile duration and frequency mismatches across pretraining and downstream tasks, and a set of cooperative objectives that jointly capture waveform morphology, physiological robustness, and long-term clinical semantics via patient-level anchoring.

Extensive experiments across four heterogeneous downstream tasks---atrial fibrillation detection, heart rate prediction, respiratory rate prediction, and heart rate estimation---demonstrate that CAP learns clinically grounded and transferable universal PPG representations, consistently outperforming strong baselines. In particular, CAP achieves an \textbf{88\%} relative improvement in MAE on respiratory rate prediction and an average relative improvement of \textbf{26.7\%} across all four tasks. Overall, our results suggest that incorporating long-term patient context and clinical semantics is a promising direction toward universal foundation models for physiological signals.

We further conduct extensive interpretability analyses to better understand CAP and validate key design choices. Ablation studies confirm the effectiveness of each component and the rationale behind major hyperparameter settings. In addition, we analyze the learned representations from multiple perspectives, including attention-based signal attribution derived from encoder weights, text--signal alignment behavior, and UMAP visualizations of the representation space. These results provide converging evidence for the effectiveness of clinical anchoring and improve the transparency of our approach.

\printbibliography

\appendix

\section{GenAI Disclosure}
We clarify the use of GenAI during the preparation of this manuscript. 
GenAI were employed exclusively for supportive purposes in three aspects:  
(i) assisting in the translation of early drafts written in the authors’ native language,  
(ii) providing linguistic refinement such as correcting grammar, improving sentence structure, and enhancing readability, and  
(iii) offering technical assistance in code editing and debugging (e.g., clarifying library usage, resolving implementation issues, or refactoring scripts for clarity).  

It is important to emphasize that all core scientific ideas, methodological contributions, analyses, and results presented in this work are entirely the intellectual product of the human authors. GenAI functioned strictly as auxiliary tools for writing and coding support, and were not involved in the conceptual or analytical aspects of the research.

\section{Limitations and Ethical Considerations}
Limitations: our datasets are predominantly from Western populations and may not generalize globally; patient-level supervision requires multiple samples per patient which may not be available in all clinical scenarios.

Ethics: all data is de-identified; the method is intended for research and wellness monitoring, not diagnostic use without FDA/clinical approval; we have carefully considered privacy implications of patient-level learning and ensure no individual patient can be re-identified from the learned representations.

\begin{table}[h]
\centering
\caption{Statistics of the Pre-training Dataset Before and After Curation}
\label{tab:appendix_dataset_stats}
\small
\resizebox{\columnwidth}{!}{%
\begin{tabular}{lcc}
\toprule
\textbf{Category / Metric} & \textbf{Source Dataset} & \textbf{Processed Dataset} \\
\midrule
\multicolumn{3}{l}{\textit{Dataset Scale}} \\[0.5ex] 
Total Samples (Segments/Pairs) & 721,840 & 31,393 \\
Total Patients & 52,398 & 2,279 \\
Total Effective Beats & 278,551,511 & 11,118,996 \\
\midrule
\multicolumn{3}{l}{\textit{Patient Window Distribution}} \\[0.5ex]
Mean Windows per Patient & 13.78 & 15.32 \\
Range of Windows per Patient & 1--8,244 & 5--8,244 \\
\midrule
\multicolumn{3}{l}{\textit{Signal Quality Profiles}} \\[0.5ex]
Mean SQI Score & 88.29\% & 93.47\% \\
Samples with Excellent Quality ($>90\%$) & 400,252 (55.5\%) & 31,393 (100.0\%) \\
Samples with Good Quality ($>70\%$) & 675,486 (93.7\%) & 31,393 (100.0\%) \\
\bottomrule
\end{tabular}
}
\end{table}

\section{Evaluation Protocols and Fairness.} 
Delivering strictly controlled comparisons within the PPG foundation model community remains an open challenge, primarily due to fragmented evaluation protocols and the lack of publicly available source code or checkpoints in prior arts (e.g., GPT-PPG, SiamQuality). To guarantee a rigorous and fair evaluation, we establish absolute consistency with the pipeline of GPT-PPG—the direct predecessor of our work—strictly replicating its dataset splits, preprocessing procedures, and evaluation metrics. Furthermore, unlike several baselines that rely on quality-based signal filtering, we evaluate our model on 100\% of the raw data. This represents a more challenging and realistic setting, and we explicitly delineate the varying protocol configurations of each baseline in our comprehensive results.

\section{Data Preprocessing and Signal Quality Specifications}
\label{appendix:sqi_details}

To guarantee the fidelity of the multi-modal representation alignment during pre-training, we employ a rigorous artifact filtering pipeline based on the Signal Quality Index (SQI). We utilize the open-source \texttt{pyPPG} library to compute multi-dimensional quality metrics, which comprehensively evaluate signal morphology, periodicity, and amplitude characteristics of each photoplethysmography (PPG) segment. 

In clinical settings—particularly within the Emergency Department (ED)—finger-pinch PPG acquisition is highly susceptible to severe noise. The dominant noise sources in our cohort include equipment contact instability (e.g., probe displacement) and patient motion artifacts during acute care. To mitigate these confounding factors, we enforce a stringent inclusion threshold of $\text{SQI} > 90\%$. This strict filtering, combined with the prerequisite of non-empty matched clinical text (Electronic Health Records), yields a high-fidelity pre-training corpus consisting of 2,279 unique patient pairs.

Table~\ref{tab:appendix_dataset_stats} provides a comprehensive breakdown of the data characteristics, delineating the distribution and quality metrics before and after the inclusion and exclusion criteria.

\end{document}